\documentclass[twocolumn,superscriptaddress,amsmath,amssymb,pra,longbibliography]{revtex4-1}
\usepackage{amssymb,amsmath}
\usepackage{cmap}
\usepackage{graphicx}
\usepackage{bm}% bold math
\usepackage{color}
\usepackage{blindtext}
\usepackage{hyperref}
\usepackage{listings}
\usepackage{booktabs}
\usepackage{multirow}
\usepackage{amsmath}
\usepackage{mathrsfs}
\usepackage{braket}
\usepackage{mathtools}

\usepackage{dsfont}
\usepackage{tabularx}

\newcommand{\red}[1]{{\color{black}{#1}}}

\begin{document}

\title{Momentum, angular momentum, and spin of waves in \red{an isotropic collisionless}  plasma}

\author{Konstantin Y. Bliokh}
\affiliation{Theoretical Quantum Physics Laboratory, RIKEN Cluster for Pioneering Research, Wako-shi, Saitama 351-0198, Japan}
\author{Yury P. Bliokh}
\affiliation{Physics Department, Technion, Israel Institute of Technology, Haifa 320003, Israel}

%\date{\today}

\begin{abstract}
We examine the momentum and angular momentum (including spin) properties of linear waves, both longitudinal (Langmuir) and transverse (electromagnetic), in an isotropic nonrelativistic collisionless electron plasma. We focus on conserved quantities associated with the translational and rotational invariance of the wave fields with respect to the homogeneous medium; these are sometimes called pseudo-momenta. There are two types of the momentum and angular momentum densities: (i) the kinetic ones associated with the energy flux density and the symmetrized (Belinfante) energy-momentum tensor and (ii) the canonical ones associated with the conserved Noether currents and canonical energy-momentum tensor. We find that the canonical momentum and spin densities of Langmuir waves are similar to those of sound waves in fluids or gases; they are expressed via the electron velocity field. In turn, the momentum and spin densities of electromagnetic waves can be written either in the forms known for free-space electromagnetic fields, involving only the electric field, or in the dual-symmetric forms involving both electric and magnetic fields, as well as the effective permittivity of plasma. We derive these properties both within the phenomenological macroscopic approach and microscopic Lagrangian field theory for the coupled electromagnetic fields and electrons. Finally, we explore implications of the canonical momentum and spin densities in transport and electrodynamic phenomena: the Stokes drift, the wave-induced magnetization (inverse Faraday effect), etc.
\end{abstract}

%\keywords{Acoustic force; canonical momentum; acoustic spin; acoustic torque.}

\maketitle

%%%%%%%%%%%%%%%%%%%%%%%%%%%%%%
\section{Introduction}
%%%%%%%%%%%%%%%%%%%%%%%%%%%%%%

Determination of fundamental conserved quantities, such as energy, momentum, and angular momentum, is an important problem in all areas of wave physics, both classical and quantum. Despite the presence of well developed theoretical methods, such as Lagrangian field theory and Noether's theorem, this problem is by no means trivial. One can mention longstanding discussions on the (angular) momentum and pseudomomentum of acoustic, elastic, and water waves \cite{McIntyre1981,Peierls_I,Peierls_II,Stone2002}, canonical and kinetic (angular) momentum in relativistic field theory \cite{Belinfante1940,Rosenfeld1940,Soper,Leader2014}, the Abraham-Minkowski dilemma in optics \cite{Brevik1979,Pfeifer2007,Barnett2010_II,Milonni2010,Kemp2011}, etc. Notably, the wave momentum identification is often closely related to the proper description of {\it spin}, i.e., the intrinsic angular momentum.   

Recently, there were numerous attempts to conquer some of these problems with new theoretical and experimental approaches, including proper consideration of spin of elastic \cite{Zhang2014,Nakane2018,Long2018,Streib2021}, sound \cite{Jones1973,Shi2019,Toftul2019,Burns2020}, and water \cite{Bliokh2022} waves, as well as revisiting definitions of the momentum and spin densities for optical fields in free space \cite{Berry2009,Barnett2010,Cameron2012,Bliokh2013NJP,Cameron2013,Bliokh2014NC,Bliokh2014NJP,Bliokh2015PR,Aiello2015} and in media \cite{Bliokh2017PRL,Bliokh2017NJP,Picardi2018}. It should be noted that, while many of earlier approaches discussed the (angular) momentum of the simplest {\it plane} waves with well-defined wavevector and frequency, recent approaches usually consider more general {\it structured} (inhomogeneous) waves, which can consist of many interfering plane waves (yet often restricted by the monochromaticity) \cite{Rubinsztein_Dunlop_2016}.  

In this work, using recent achievements in the description of the momentum and angular momentum of structured electromagnetic and acoustic waves, we examine the momentum, angular momentum, and spin of waves in an {ideal} nonrelativistic electron plasma {within the hydrodynamical model} \cite{Krall,Akhiezer_book}. This problem is important by itself for general theory of plasma waves and for applications to structured plasma waves \cite{Esarey2009,Mendonca2009,Vieira2018,Blackman2019}, as well as a model for the general approach to the wave momentum and angular momentum in complex systems \cite{Dewar1977,Dodin2012}. Waves in plasma have remarkable peculiarities which distinguish this system from simpler sound or optical waves. First, this is a coupled light-matter system which allows both macroscopic (in terms of an effective medium) and microscopic (in terms of individual electrons and fields) descriptions. Second, an isotropic homogeneous electron plasma possesses two types of modes: longitudinal (akin to sound waves) and transverse (electromagnetic), which have different velocities and dispersions. This provides a perfect platform for applications of the general theory of wave momentum and angular momentum.

\red{In addition, electromagnetic waves in an electron plasma have important implications in plasmonics, i.e., optics involving metals \cite{Maier_book,Zayats2005}. The Drude model of metals is essentially equivalent to the electron plasma considered in this work, whereas the momentum and angular momentum properties of electromagnetic waves in metals, including surface plasmon-polaritons, are highly important for modern nanooptics \cite{Bliokh2017PRL,Bliokh2017NJP,Bliokh2015NP}.}

Note that different types of momenta in different areas of wave physics lack for universal terminology. The terms ``pseudomomentum'', ``canonical momentum'', ``wave momentum'' can have different physical meanings depending on the field and context. To be specific, we consider momentum and angular momentum related to the translational and rotational symmetries of the wave fields with respect to the homogeneous medium (plasma). In fluid mechanics and acoustics (and sometimes in the Abraham-Minkowski discussions), these quantities are often called `{\it pseudomomentum}' and `{\it angular pseudomomentum}' \cite{McIntyre1981,Peierls_I,Peierls_II,Stone2002,Nakane2018,Streib2021,Gordon1973,Nelson1991} (in contrast to quantities associated with translations and rotations of the whole system ``waves + medium''). In addition, these quantities allow different {\it kinetic} and {\it canonical} forms, associated with the canonical and Belinfante (symmetrized) forms of the energy-momentum tensor \cite{Belinfante1940,Rosenfeld1940,Soper,Leader2014,Bliokh2013NJP,Bliokh2014NC,Bliokh2014NJP,Burns2020, Bliokh2022}. \red{While the kinetic quantities have more universal forms (like, e.g., the Poynting vector), not restricted by the monochromaticity,  the canonical momentum and angular momentum densities are directly related to observable phenomena, mechanical motion of electrons in plasma, and also involve the spin density as an independent quantity. }

We derive both the kinetic and canonical (pseudo-)momentum and angular (pseudo-)momentum densities for longitudinal (Langmuir) and transverse (electromagnetic) waves in plasma. In doing so, we use both phenomenological macroscopic approach and microscopic derivation based on the Lagrangian field theory for the coupled ``fields + electrons'' system. We show that the kinetic momentum density is given by the energy flux density divided by the product of the phase and group velocities. 

In turn, the canonical momentum and spin densities acquire simple meaningful forms in the case of monochromatic fields and time averaging of these quantities.
For Langmuir waves, the canonical momentum and spin densities are entirely similar to those in sound waves and are explicitly expressed via the electron velocity field. For electromagnetic waves, the situation is more sophisticated, because the canonical momentum and spin densities can be written in the standard (involving only the electric field) and dual-symmetric (involving both electric and magnetic fields) forms. While the dual-symmetric form looks appealing \cite{Berry2009,Barnett2010,Cameron2012,Bliokh2013NJP,Cameron2013,Bliokh2014NC,Bliokh2014NJP,Bliokh2015PR,Aiello2015,Bliokh2017PRL,Bliokh2017NJP,Dressel2015,Burns2020}, the microscopic Lagrangian formalism for the coupled fields-electron system results in the momentum and spin densities explicitly involving only the electric field and no particle characteristics. 

Remarkably, \red{we find that the Abraham-Minkowski dilemma, i.e., the main problem for the electromagnetic wave momentum in a medium, does not appear for electromagnetic waves in an ideal electron plasma}, because the Abraham and Minkowski (with the proper dispersive corrections \cite{Nelson1991,Philbin2011,Philbin2012,Philbin2012_II,Bliokh2017NJP,Bliokh2017PRL}) plane-wave momenta become equal to each other. In the interpretation of Ref.~\cite{Partanen2017}, this is associated with the fact that electromagnetic waves in plasma do not excite mass density waves {in the approximation under consideration}.

Finally, we consider interesting implications of the canonical momentum and spin densities in plasma waves: the {\it Stokes drift} of the electrons \cite{Stokes,Bremer2017,Bliokh2022} and the associated direct electric current, as well as the microscopic rotation of the electrons and the appearance of the {\it magnetization} of plasma (inverse Faraday effect) \cite{Bezzerides1977,Mora1978,Pitaevskii1961,LL_ED,Hertel2006,Bliokh2017PRL,Bliokh2017NJP}.

%%%%%%%%%%%%%%%%%%%%%%%%%%%%%%
\section{General equations}
%%%%%%%%%%%%%%%%%%%%%%%%%%%%%%

Linear waves in a nonrelativistic homogeneous collisionless electron plasma are described, {within the hydrodynamical model}, by the equations of motion for electrons (including the continuity equation) coupled to Maxwell's equations for electromagnetic field \cite{Krall,Akhiezer_book}:  
\begin{eqnarray}
\label{eq1}
 m && \frac{\partial {\bf v}}{\partial t} = e\, {\bf E} - \beta\, {\bm \nabla} n \, , \nonumber \\
&& \frac{\partial n}{\partial t} = - n_0 {\bm \nabla} \cdot {\bf v}  \, , \nonumber \\
&& \frac{\partial {\bf E}}{\partial t} = c\, {\bm \nabla}\times {\bf H} - 4\pi e n_0 {\bf v}  \, ,  \nonumber \\
&& \frac{\partial {\bf H}}{\partial t} = - c\, {\bm \nabla}\times {\bf E} \, , \quad
{\bm \nabla} \cdot {\bf H} =0 \, .
\end{eqnarray}
Here, $m$ and $e<0$ are the electron mass and charge, $n_0$ is the unperturbed density of electrons, $c$ is the speed of light, ${\bf v}$ is the electron velocity, $n$ is the perturbation of the electron density, $\beta = 3k_B T/n_0$ (where $k_B$ is the Boltzmann constant and $T$ is the electron-gas temperature), and we use the Gaussian units. Note that the forth Maxwell equation ${\bm \nabla} \cdot {\bf E} = 4\pi e n$ is not independent and follows from Eqs.~(\ref{eq1}). {The heavy ion background neutralizes the total electric charge of the electrons and is assumed to be motionless in this approximation.}

Equations (\ref{eq1}) obey the conservation law
\begin{equation}
\label{eq2}
\frac{\partial W}{\partial t} + {\bm \nabla}\cdot {\bf U} = 0 \, ,
\end{equation}
where 
\begin{equation}
\label{eq3}
W = mn_0\frac{{\bf v}^2}{2} + \beta\frac{n^2}{2} + \frac{1}{8\pi}\!\left({\bf E}^2 +{\bf H}^2\right)
\end{equation}
can be associated with the wave energy density (kinetic + potential + electromagnetic-field parts), whereas
\begin{equation}
\label{eq4}
{\bf U} = \beta n_0 n {\bf v} + \frac{c}{4\pi} ({\bf E}\times {\bf H})
\end{equation}
is the energy flux density. Note that the electron and field parts of equations  (\ref{eq3}) and (\ref{eq4}) are entirely similar to the energy density and the energy flux density for sound waves in a neutral gas and for electromagnetic waves in the vacuum \cite{LLfluid,LLfield}.

Equations~(\ref{eq1}) describe two types of electron plasma waves: (i) longitudinal Langmuir waves and (ii) transverse electromagnetic waves \cite{Krall,Akhiezer_book}. The Langmuir waves are characterized by the relations and dispersion as follows: 
\begin{equation}
\label{eq5}
{\bf H} = {\bm \nabla}\times {\bf E} = {\bm \nabla}\times {\bf v} = {\bf 0}\,, ~~~\omega^2 = \omega_p^2 + \frac{\beta n_0}{m}k^2\,,
\end{equation}
where $\omega_p = \sqrt{4\pi n_0 e^2/m}$ is the plasma frequency. Note that the group and phase velocities of these waves satisfy $v_g v_p =\beta n_0/m$. In turn, the electromagnetic waves are characterized by the relations 
\begin{equation}
\label{eq6}
n = {\bm \nabla}\cdot {\bf E} = {\bm \nabla} \cdot {\bf v} = {0}\,, \quad \omega^2 = \omega_p^2 +c^2k^2\,.
\end{equation}
Their group and phase velocities satisfy $v_g v_p =c^2$. 
Below we analyze the momentum and angular momentum properties of these two types of plasma waves.

Since the canonical momentum and angular momentum (including orbital and spin parts) of waves acquires particularly clear and meaningful form for monochromatic waves \cite{Berry2009,Bliokh2013NJP,Bliokh2014NC,Bliokh2015PR,Aiello2015,Bliokh2017PRL,Bliokh2017NJP,Bliokh2014NJP,Shi2019,Toftul2019,Burns2020,Bliokh2022}, we will use complex monochromatic field amplitudes indicated by the overbar: ${\bf v}({\bf r},t) = {\rm Re}\!\left[ \bar{\bf v}({\bf r}) e^{-i\omega t} \right]$, $n({\bf r},t) = {\rm Re}\!\left[ \bar{n}({\bf r}) e^{-i\omega t} \right]$, etc. The complex amplitudes satisfy Eqs.~(\ref{eq1}) with $\partial/\partial t \to -i\omega$. Quadratic forms (energy, momentum, angular momentum, etc.) for monochromatic fields are averaged over the period of oscillations and marked with the overbar. For instance, the time-averaged monochromatic version of Eq.~(\ref{eq4}) is $\bar{\bf U} = \dfrac{\beta n_0}{2}\, {\rm Re}\!\left(\bar{n}^* \bar{\bf v}\right) + \dfrac{c}{8\pi}\,{\rm Re}\!\left(\bar{\bf E}^*\!\times \bar{\bf H}\right)$.

As we will see, the {\it kinetic momentum} density and the corresponding {\it kinetic angular momentum} density of a wave can be defined as
\begin{equation}
\label{eq7}
{\bm{\mathcal P}} = \frac{\bf U}{v_g v_p} \, , \quad
{\bm{\mathcal J}} = {\bf r} \times {\bm{\mathcal P}} \, .
\end{equation}
For monochromatic waves, the time-averaged kinetic momentum density can be decomposed into canonical and spin parts according to the general Belinfante-Rosenfeld field-theory relation \cite{Belinfante1940,Rosenfeld1940,Soper,Leader2014,Berry2009,Bliokh2013NJP,Bliokh2014NC,Bliokh2015PR, Aiello2015,Shi2019,Burns2020,Bliokh2022}:
\begin{equation}
\label{eq8}
\bar{\bm{\mathcal P}} = \bar{\bf P} + \frac{1}{2} {\bm \nabla}\times \bar{\bf S} \, .
\end{equation}
Here, $ \bar{\bf P}$ is the {\it canonical momentum} density and $\bar{\bf S}$ is the {\it spin} density. The corresponding canonical angular momentum density consists of the orbital (extrinsic) and spin (intrinsic) parts:  
\begin{equation}
\label{eq9}
\bar{\bf J} = {\bf r} \times \bar{\bf P} + \bar{\bf S} \, .
\end{equation}
Notably, the canonical momentum and angular momentum densities can be derived directly from the field Lagrangian and Noether's theorem \cite{Soper,Leader2014,Bliokh2013NJP,Dressel2015,Nakane2018,Burns2020,Streib2021}. These quantities provide the generators of translations and rotations of the wave field with respect to the homogeneous medium, i.e., these are `pseudo'-momenta \cite{McIntyre1981,Peierls_I,Peierls_II,Stone2002,Nakane2018,Streib2021,Gordon1973,Nelson1991}. Furthermore, the canonical momentum and spin densities are often more meaningful and directly observable in experiments; this has been demonstrated in optics \cite{Huard1978,Matsudo1998,ONeil2002,Garces2003,Barnett2013,Bliokh2014NC,Bliokh2015PR,Aiello2015,Neugebauer2015}, acoustics \cite{Shi2019,Toftul2019}, and hydrodynamics \cite{Bliokh2022}. 
%In elastic solids and fluids, the canonical quantities are sometimes called `pseudomomentum' and `angular pseudomomentum' . 

%%%%%%%%%%%%%%%%%%%%%%%%%%%%%%
\section{Langmuir waves}
%%%%%%%%%%%%%%%%%%%%%%%%%%%%%%

For Langmuir waves (\ref{eq5}), the energy density (\ref{eq3}) and energy flux density (\ref{eq4}) are reduced to 
\begin{equation}
\label{eq10}
W_L = mn_0\frac{{\bf v}^2}{2} + \beta\frac{n^2}{2} + \frac{{\bf E}^2}{8\pi} \, , \quad
{\bf U}_L = \beta n_0 n {\bf v} \, .
\end{equation}
We first apply Eqs.~(\ref{eq7})--(\ref{eq9}) phenomenologically, and then show how the momentum and angular momentum wave properties can be derived within the rigorous Lagrangian field-theory formalism. 

The kinetic momentum density (\ref{eq7}) yields for Langmuir waves: 
\begin{equation}
\label{eq11}
{\bm{\mathcal P}}_L 
%= \frac{{\bf U}_L}{v_g v_p} 
= m n {\bf v} \, .
\end{equation}
This quantity is clearly associated with the second-order perturbation of the momentum of the electrons, which remains non-zero after time averaging in monochromatic fields: $\bar{\bm{\mathcal P}}_L = \dfrac{m}{2}\, {\rm Re}\!\left( \bar{n}^* \bar{\bf v}\right)$. Equations (\ref{eq11}) is equivalent to the kinetic momentum density of sound waves in gases or fluids \cite{LLfluid,Shi2019,Burns2020,Bliokh2022}.

Using Eqs.~(\ref{eq1}) and (\ref{eq5}), the kinetic momentum density (\ref{eq11}) can be written in the Belinfante-Rosenfeld form (\ref{eq8}) with the canonical momentum and spin densities
\begin{eqnarray}
\label{eq12}
\bar{\bf P}_L & = & \frac{mn_0}{2\omega}\, {\rm Im}\!\left[ \bar{\bf v}^*\!\cdot ({\bm \nabla}) \bar{\bf v}\right] , \\ [5pt]
\label{eq13}
\bar{\bf S}_L & = & \frac{mn_0}{2\omega}\, {\rm Im}\!\left( \bar{\bf v}^*\! \times \bar{\bf v}\right) ,
\end{eqnarray}
\red{where we used the notation ${\bf X}\cdot ({\bm \nabla}) {\bf Y} = \sum_{i=1}^{3}  X_i {\bm \nabla} Y_i$ \cite{Berry2009}.} These densities are also equivalent to their sound-wave counterparts \cite{Shi2019,Burns2020,Bliokh2022}. 

It is important to emphasize that the canonical momentum (\ref{eq12}) and spin (\ref{eq13}) have clear mechanical interpretations. Namely, the spin density represents mechanical angular momentum density for electrons following microscopic first-order elliptical trajectories in generic inhomogeneous wave fields. This is seen by time-averaging the mechanical angular momentum density of the electron, $mn_0 ({\bf a} \times {\bf v})$, where ${\bf a}$ is the electron displacement, ${\bf v} = \partial {\bf a}/\partial t$ or $\bar{\bf v} = -i\omega \bar{\bf a}$ \cite{Nakane2018,Jones1973,Shi2019,Bliokh2017NJP,Bliokh2022}. In turn, the canonical momentum density can be associated with the {\it Stokes drift} phenomenon known in fluid mechanics \cite{Stokes,Bremer2017,Bliokh2022}. It is the second-order drift effect related to the difference between the Lagrangian and Eulerian velocities of the particle in a gas or fluid. One can show (see Section V) that the canonical momentum density (\ref{eq12}) is related to the Stokes drift velocity ${\bf u}$ as $\bar{\bf P}_L = m n_0 {\bf u}_L$.

For a plane wave, there is no difference between the kinetic and canonical momentum, and Eqs.~(\ref{eq10})--(\ref{eq13}) yield
\begin{eqnarray}
\label{PW_L}
\bar{\bm{\mathcal P}}_{L} = \bar{\bf P}_{L} =\frac{\bar{W}_{L}}{\omega}\, {\bf k}\,, \quad 
\bar{\bf S}_{L} = {\bf 0}\,,
\end{eqnarray}
which agrees with Ref.~\cite{Dodin2012}. Here the spin vanishes because of the longitudinal character of Langmuir waves. However, the spin density (\ref{eq13}) is generally nonzero for an inhomogeneous interference field where the local polarization is generally elliptic: to observe this, it is sufficient to interfere two plane waves with non-collinear wavevectors \cite{Bekshaev2015,Bliokh2015PR,Shi2019,Bliokh2022}. Note that for longitudinal waves, one can represent the vector velocity field via scalar potential: ${\bf v} = {\bm \nabla}\phi$. Then, the spin density (\ref{eq13}) is a solenoidal field $\propto {\rm Im}\!\left( {\bm \nabla}\bar{\phi}^*\! \times {\bm \nabla}\bar{\phi}\right) = {\bm \nabla} \times {\rm Im}\!\left( \bar{\phi}^* {\bm \nabla}\bar{\phi}\right)$, and its integral over the space vanishes for a localized wavefield \cite{Bliokh2019_II}. This means that, akin to sound waves, Langmuir waves are globally spinless, but generically have a nonzero local spin density.

Thus, the momentum and angular-momentum properties of Langmuir waves are entirely similar to those of longitudinal sound waves and do not depend explicitly on the electric field ${\bf E}$. Apparently, this is because the electric field is purely longitudinal in such waves, ${\bm \nabla}\times {\bf E} = {\bf 0}$, and does not carry any momentum or angular momentum (although it can be locally rotating in a generic inhomogeneous wave field). 

To derive the momentum and angular momentum properties of Langmuir waves in a more rigorous way, we introduce the Lagrangian density including the kinetic and potential energy of the electron fluid, electric field energy, and the minimal electron-field coupling:
\begin{equation}
\label{eq14}
{\cal L}_L = \underbracket{\frac{mn_0}{2} \dot{\bf a}^2\!}_\text{kinetic}  - 
\underbracket{\frac{\beta n_0^2}{2}\!\left( {\bm \nabla} \cdot {\bf a}\right)^2}_\text{potential}\! + \underbracket{\frac{1}{8\pi} \dot{\bf A}^2}_\text{field}\! + \underbracket{en_0 \dot{\bf a}\cdot {\bf A}}_\text{coupling}\,.
\end{equation}
Here $\dot{...} \equiv \partial ... / \partial t$, whereas the potentials ${\bf a}$ and ${\bf A}$ are related to the wave fields as ${\bf v} = \dot{\bf a}$, $n = - n_0 {\bm \nabla} \cdot {\bf a}$, and ${\bf E} = - \dot{\bf A}$. Obviously, ${\bf a}$ is the electron displacement, whereas ${\bf A}$ is the electromagnetic vector-potential in the {Weyl} gauge {(up to the factor of $c$)}. The Euler-Lagrange equations for the Lagrangian density (\ref{eq14}) yield the first and the third of the equations of motion (\ref{eq1}); the second (continuity) equation follows from the representation of ${\bf v}$ and $n$ via the potential ${\bf a}$.   

The canonical energy-momentum tensor $T^{\mu\nu}$ can be derived from the Lagrangian (\ref{eq14}) and Noether's theorem with respect to the space-time translation symmetries ${\bf r} \to {\bf r} + \delta{\bf r}$, $t \to t + \delta t$ \cite{Soper,LLfield}. This yields the energy density and the energy flux density (\ref{eq10}):
\begin{eqnarray}
&&T^{00}_L= \frac{\partial {\cal L}_L}{\partial \{ \dot{\bf a},\dot{\bf A}\}} \cdot \{ \dot{\bf a},\dot{\bf A}\} -  {\cal L}_L = W_L\,, \nonumber \\
\label{eq15}
&&T^{0i}_L= \frac{\partial {\cal L}_L}{\partial \{ \nabla_i{\bf a},\nabla_i{\bf A}\}} \cdot \{ \dot{\bf a},\dot{\bf A}\} = U_{Li}\,,
\end{eqnarray}
as well as the canonical momentum density:
\begin{eqnarray}
\label{eq16}
&&T^{i0}_L= - \frac{\partial {\cal L}_L}{\partial \{ \dot{\bf a},\dot{\bf A}\}} \cdot (\nabla_i) \{{\bf a},{\bf A}\} = P_i\,, \\
&&{\bf P}_L = -mn_0 {\bf v}\cdot ({\bm \nabla}){\bf a} +\frac{1}{4\pi}{\bf E}\cdot ({\bm \nabla}){\bf A} - en_0 {\bf A}\cdot ({\bm \nabla}){\bf a}\,. \nonumber 
\end{eqnarray}
Performing the time averaging of the canonical momentum density (\ref{eq16}) for monochromatic fields with 
$\bar{\bf a} = i\omega^{-1} \bar{\bf v}$ and $\bar{\bf A} = -i\omega^{-1} \bar{\bf E} = - (4\pi en_0/\omega^2) \bar{\bf v}$, we find that the last two terms in Eq.~(\ref{eq16}), originating from the field and coupling terms in the Lagrangian (\ref{eq14}), cancel each other, while the remaining first (kinetic) term yields Eq.~(\ref{eq12}).

Finally, the canonical angular momentum of Langmuir waves is also derived from the Lagrangian density (\ref{eq14}) and Noether theorem with respect to spatial rotations \cite{Soper,Bliokh2013NJP,Nakane2018}. This results in the angular momentum density 
\begin{eqnarray}
\label{eq17}
{\bf J}_L\, && = - \frac{\partial {\cal L}_L}{\partial \{ \dot{\bf a},\dot{\bf A}\}} \cdot ({\bf r} \times {\bm \nabla}) \{{\bf a},{\bf A}\} -
\frac{\partial {\cal L}_L}{\partial \{ \dot{\bf a},\dot{\bf A}\}} \times \{{\bf a},{\bf A}\} \nonumber \\
&&\equiv {\bf r} \times {\bf P}_L +{\bf S}_L\,, 
\end{eqnarray}
where the spin part becomes
\begin{equation}
\label{eq18}
{\bf S}_L = -mn_0 {\bf v}\times{\bf a} +\frac{1}{4\pi}{\bf E}\times{\bf A} - en_0 {\bf A}\times{\bf a}\,. 
\end{equation}
Similarly to the canonical momentum density (\ref{eq16}), the last two terms (field and coupling) in Eq.~(\ref{eq18}) are mutually cancelled after the time averaging in monochromatic fields, while the first (kinetic) term yields Eq.~(\ref{eq13}).

%%%%%%%%%%%%%%%%%%%%%%%%%%%%%%
\section{Electromagnetic waves}
%%%%%%%%%%%%%%%%%%%%%%%%%%%%%%
 
For transverse electromagnetic waves (\ref{eq6}), the energy density (\ref{eq3}) and the energy flux density (\ref{eq4}) read:
\begin{equation}
\label{eq19}
W_{EM} = mn_0\frac{{\bf v}^2}{2}  + \frac{1}{8\pi}\!\left({\bf E}^2 + {\bf H}^2\right) , ~~
{\bf U}_{EM} = \frac{c}{4\pi} ({\bf E}\times {\bf H}) \, .
\end{equation}
The energy flux density is the electromagnetic Poynting vector \cite{LLfield}, whereas the energy density is a sum of the kinetic energy of the electrons and the electromagnetic field energy. For monochromatic fields, using $\bar{\bf v}=(ie/m\omega) \bar{\bf E}$, the time-averaged energy density can be written as 
\begin{equation}
\label{eq20}
\bar{W}_{EM} = \frac{1}{16\pi}\!\left(\tilde{\varepsilon}\,|\bar{\bf E}|^2 + |\bar{\bf H}|^2\right) , 
\end{equation}
where $\varepsilon = 1- \omega_p^2/\omega^2$ is the permittivity of plasma, and $\tilde \varepsilon = d(\omega \varepsilon)/d\omega = 1+ \omega_p^2/\omega^2$. Equation (\ref{eq20}) is the Brillouin energy density for electromagnetic waves in a dispersive non-magnetic medium \cite{LL_ED}

Like in the case of Langmuir waves, we first apply general Eqs.~(\ref{eq7})--(\ref{eq9}) phenomenologically. 
The kinetic momentum density (\ref{eq7}) yields 
\begin{equation}
\label{eq21}
{\bm{\mathcal P}}_{EM} 
%= \frac{{\bf U}_{EM}}{v_g v_p} 
= \frac{1}{4\pi c} ({\bf E}\times {\bf H}) \, .
\end{equation}
This quantity is known as the {\it Abraham momentum} density for electromagnetic waves in a medium \cite{Brevik1979,Pfeifer2007,Barnett2010_II,Milonni2010,Kemp2011}. 

Using Eqs.~(\ref{eq1}) and (\ref{eq6}), we decompose the kinetic momentum density (\ref{eq21}) into the Belinfante-Rosenfeld form (\ref{eq8}) with the canonical momentum and spin parts. Importantly, this decomposition is not unique; it depends on the choice of either electric, or magnetic, or both fields as the main source for these quantities \cite{Berry2009,Barnett2010,Bliokh2013NJP}. Below we consider two most important options for the canonical momentum and spin: (i) the `standard' one based on the electric field and (ii) the `dual-symmetric' one keeping the dual symmetry between the electric and magnetic fields.

For the first option, we obtain
\begin{eqnarray}
\label{eq22}
\bar{\bf P}_{E} & = & \frac{1}{8\pi\omega}\, {\rm Im}\!\left[ \bar{\bf E}^*\!\cdot ({\bm \nabla}) \bar{\bf E}\right]  , \\ [5pt]
\label{eq23}
\bar{\bf S}_{E} & = & \frac{1}{8\pi\omega}\, {\rm Im}\!\left( \bar{\bf E}^*\! \times \bar{\bf E}\right) .
\end{eqnarray}
These densities have the forms of the electric-field-based canonical momentum and spin densities for electromagnetic waves in free space \cite{Soper,Bliokh2013NJP}. Note that, in contrast to Langmuir waves, the spin (\ref{eq23}) of electromagnetic waves does not vanish even for a plane wave, where it is associated with the degree of circular polarization (helicity) and directed along the wavevector \cite{Bliokh2015PR}. 

Mechanical properties of the electrons do not appear explicitly in Eqs.~(\ref{eq22}) and (\ref{eq23}), although these properties must contribute to the momentum and angular momentum of electromagnetic waves in plasma. 
Therefore, we suggest the decomposition of Eqs.~(\ref{eq22}) and (\ref{eq23}) into the mechanical and electrodynamic (field) parts as follows:
\begin{eqnarray}
\label{eq24}
\bar{\bf P}_E^{\rm m}\! & = & \frac{mn_0}{2\omega}\, {\rm Im}\!\left[ \bar{\bf v}^*\!\cdot ({\bm \nabla}) \bar{\bf v}\right]= \frac{1}{16\pi}\frac{d\varepsilon}{d\omega}\, {\rm Im}\!\left[ \bar{\bf E}^*\!\cdot ({\bm \nabla}) \bar{\bf E}\right]\! , \\ [5pt]
\label{eq25}
\bar{\bf S}_E^{\rm m}\! & = & \frac{mn_0}{2\omega}\, {\rm Im}\!\left( \bar{\bf v}^*\!\times \bar{\bf v}\right)= \frac{1}{16\pi}\frac{d\varepsilon}{d\omega}\, {\rm Im}\!\left( \bar{\bf E}^*\! \times \bar{\bf E}\right) , \\ [5pt]
\label{eq26}
\bar{\bf P}_{E}^{\rm f}\! & = & \frac{\varepsilon}{8\pi\omega}\, {\rm Im}\!\left[ \bar{\bf E}^*\!\cdot ({\bm \nabla}) \bar{\bf E}\right]  , \\ [5pt]
\label{eq27}
\bar{\bf S}_{E}^{\rm f}\! & = & \frac{\varepsilon}{8\pi\omega}\, {\rm Im}\!\left( \bar{\bf E}^*\! \times \bar{\bf E}\right) .
\end{eqnarray}
Here we used  relations $\bar{\bf v}=(ie/m\omega) \bar{\bf E}$, $\omega \, d\varepsilon/d\omega = 2\omega_p^2/\omega^2 = 2(1-\varepsilon)$, and chose the mechanical parts in the form of Eqs.~(\ref{eq12}) and (\ref{eq13}) because the mechanical angular momentum density from the microscopic motion of electrons still equals $mn_0 ({\bf a} \times {\bf v})$ yielding Eq.~(\ref{eq13}). 
%[association of the mechanical momentum (\ref{eq24}) with the Stokes drift is more problematic, see Section V]. 

Notably, the field parts (\ref{eq26}) and (\ref{eq27}) have the forms of the `{\it Minkowski-type}' quantities in a medium, modified with the permittivity factor $\varepsilon$, while the mechanical parts (\ref{eq24}) and (\ref{eq25}) provide the dispersive corrections to these quantities \cite{Nelson1991,Philbin2011,Philbin2012,Philbin2012_II,Bliokh2017NJP}. In particular, Refs.~\cite{Nelson1991,Bliokh2017NJP} show that these dispersive corrections originate from kinematic properties of the medium particles. The fact that the multiplication by $\varepsilon$ and dispersive corrections cancel each other to yield the free-space forms (\ref{eq22}) and (\ref{eq23}) means that {\it there is no Abraham-Minkowski dilemma} \cite{Brevik1979,Pfeifer2007,Barnett2010_II,Milonni2010,Kemp2011,Nelson1991,Philbin2011,Philbin2012,Dodin2012,Partanen2017,Bliokh2017PRL,Bliokh2017NJP} for electromagnetic waves in plasma and both approaches yield the same plane-wave momentum and spin. As we show in the Lagrangian formalism below, on the microscopic level this is explained by the mutual cancelation of the field-electron coupling contributions [included in the field quantities (\ref{eq26}) and (\ref{eq27})] and mechanical kinetic contributions, so that only the pure-field contributions determine the total momentum and spin densities (\ref{eq22}) and (\ref{eq23}). In the interpretation of Ref.~\cite{Partanen2017}, this is associated with the fact that electromagnetic waves in plasma do not excite mass density waves, and the difference between the Abraham and Minkowski momenta, proportional to $v_g - c^2/v_p$ vanishes.

In the second, `dual-symmetric' approach, we seek for the Belinfante-Rosenfeld form (\ref{eq8}) with canonical momentum and spin densities containing both electric and magnetic fields. This results in
\begin{eqnarray}
\label{eq28}
\bar{\bf P}_{EM} & = & \frac{1}{16\pi\omega}\, {\rm Im}\!\left[ \tilde\varepsilon \, \bar{\bf E}^*\!\cdot ({\bm \nabla}) \bar{\bf E} + \bar{\bf H}^*\!\cdot ({\bm \nabla}) \bar{\bf H}\right]  , \\ [5pt]
\label{eq29}
\bar{\bf S}_{EM} & = & \frac{1}{16\pi\omega}\, {\rm Im}\!\left( \tilde\varepsilon \,\bar{\bf E}^*\! \times \bar{\bf E} + \bar{\bf H}^*\! \times \bar{\bf H} \right) .
\end{eqnarray}
These expressions are equivalent to the canonical Minkowski-type momentum and spin densities for electromagnetic waves in dispersive media suggested in Refs.~\cite{Bliokh2017PRL,Bliokh2017NJP,Bekshaev2018,Picardi2018}. Akin to Eqs.~(\ref{eq24})--(\ref{eq27}), one can decompose Eqs.~(\ref{eq28}) and (\ref{eq29}) into the mechanical parts equivalent to Eqs.~(\ref{eq24}) and (\ref{eq25}) and the electrodynamic parts equal to Eqs.~(\ref{eq28}) and (\ref{eq29}) with the substitution $\tilde\varepsilon \to \varepsilon$.

\red{The advantage of the `standard' definitions (\ref{eq22}) and (\ref{eq23}) is that they have extremely simple forms similar to the Langmuir-wave Eqs.~(\ref{eq12}) and (\ref{eq13}) but with the electric field ${\bf E}$ instead of the velocity ${\bf v}$. Furthermore, we show below that the canonical momentum and spin densities (\ref{eq22}) and (\ref{eq23}) directly follow from the microscopic Lagrangian formalism. The drawback of these definitions is that they do not involve explicitly the magnetic field ${\bf H}$, which can be equally important in electromagnetic waves. In particular, this means that if only the magnetic field rotates in a given point, while the electric field has a linear polarization, the spin density vanishes in this point. The dual-symmetric definitions (\ref{eq28}) and (\ref{eq29}) restore the parity between the electric and magnetic fields, contributing to the canonical momentum and spin densities on equal footing.}

It should be emphasized that differences between the definitions of the momentum and spin densities (\ref{eq21})--(\ref{eq23}), (\ref{eq28}), and (\ref{eq29}) appear only in inhomogeneous wavefields. For a plane electromagnetic wave, all these definitions, together with Eq.~(\ref{eq20}), yield [cf. Eqs.~(\ref{PW_L})]
\begin{eqnarray}
\label{PW_P}
\bar{\bm{\mathcal P}}_{EM} & = & \bar{\bf P}_{E} = \bar{\bf P}_{EM} = \frac{\bar{W}_{EM}}{\omega} {\bf k}\,, \\
\label{PW_S}
\bar{\bf S}_{E} & = & \bar{\bf S}_{EM} = \frac{\bar{W}_{EM}}{\omega} \sigma {\bm \kappa}\,,
\end{eqnarray}
where ${\bm \kappa} = {\bf k}/k$ and $\sigma = {\bm \kappa} \cdot {\rm Im}\!\left(\bar{\bf E}^*\! \times \bar{\bf E} \right)\!/ |\bar{\bf E}|^2 \in [-1,1]$ is the plane-wave helicity.

For the sake of completeness, we consider another important quantity related to the spin and momentum: the {\it helicity density}. Helicity is an independent conserved quantity related to the dual symmetry between the electric and magnetic fields in Maxwell's equations \cite{Calkin1965,Afanasiev1996,Trueba1996,Cameron2012,Bliokh2013NJP,Fernandez2013,Cameron2013,Alpeggiani2018}. It can also be regarded as a measure of the {\it chirality} of an electromagnetic field. Using general expression for the helicity density in a dispersive electromagnetic medium \cite{Alpeggiani2018}, we derive the time-averaged helicity density of electromagnetic waves in plasma:
\begin{equation}
\label{eq_H}
\bar{\mathfrak{S}} = \frac{1}{8\pi\omega \sqrt{\varepsilon}}\, {\rm Im}\! \left( {\bf H}^* \cdot {\bf E} \right) .
\end{equation}
For a plane wave, this expression yields $\bar{\mathfrak{S}} = \sigma\, \bar{W}_{EM}/\omega$.

We are now in the position to show how the above properties of electromagnetic waves are derived within the Lagrangian field-theory formalism. It is known that the choice of the canonical quantities depending on either electric, or magnetic, or both fields is directly related to the choice of the electromagnetic field Lagrangian \cite{Bliokh2013NJP,Cameron2013}. However, the dual-symmetric Lagrangian formalism is known only for free-space fields, while for coupled light-matter systems it still represents an open challenge \cite{Cameron2014,Dressel2015}. Therefore, here we present a standard Lagrangian formalism, which is known to yield the electric-field-based canonical quantities.   

The Lagrangian density can be written as
\begin{equation}
\label{eq32}
{\cal L}_{E} = \underbracket{\frac{mn_0}{2} \dot{\bf a}^2}_\text{kinetic} + \underbracket{\frac{1}{8\pi}\! \left[ \dot{\bf A}^2 - c^2 ({\bm \nabla}\times {\bf A})^2\right]}_\text{EM field} 
 + \underbracket{en_0 \dot{\bf a}\cdot {\bf A}}_\text{coupling}\,,
\end{equation}
where the electron-displacement potential ${\bf a}$ and the electromagnetic vector-potential ${\bf A}$ (in the Coulomb gauge, {up to the $c$ factor}) are related to the wave fields as ${\bf v} = \partial {\bf a}/\partial t$, ${\bf E} = - \partial {\bf A}/\partial t$, and ${\bf H} = c\, {\bm \nabla}\times {\bf A}$. The Euler-Lagrange equations for the Lagrangian density (\ref{eq32}) yield the first and the third of the equations of motion (\ref{eq1}); the last pair of Maxwell's equations follows from the representation of ${\bf E}$ and ${\bf H}$ via the potential ${\bf A}$.   

Akin to the Langmuir-wave case, we derive the canonical energy-momentum tensor from the Lagrangian (\ref{eq32}) and Noether's theorem for the space-time translations. This yields the energy and energy flux densities (\ref{eq19}): 
\begin{eqnarray}
&&T^{00}_{E}= \frac{\partial {\cal L}}{\partial \{ \dot{\bf a},\dot{\bf A}\}} \cdot \{ \dot{\bf a},\dot{\bf A}\} -  {\cal L} = W_{EM}\,, \nonumber \\
\label{eq33}
&&T^{0i}_{E}= \frac{\partial {\cal L}}{\partial \{ \nabla_i{\bf a},\nabla_i{\bf A}\}} \cdot \{ \dot{\bf a},\dot{\bf A}\} = U_{EMi}\,,
\end{eqnarray}
as well as the canonical momentum density
\begin{eqnarray}
\label{eq34}
&&T^{i0}_{E}= - \frac{\partial {\cal L}}{\partial \{ \dot{\bf a},\dot{\bf A}\}} \cdot (\nabla_i) \{{\bf a},{\bf A}\} = P_{Ei}\,, \\
&&{\bf P}_{E} = -mn_0 {\bf v}\cdot ({\bm \nabla}){\bf a} +\frac{1}{4\pi}{\bf E}\cdot ({\bm \nabla}){\bf A} - en_0 {\bf A}\cdot ({\bm \nabla}){\bf a}\,. \nonumber 
\end{eqnarray}
Performing the time averaging of the canonical momentum density (\ref{eq34}) for monochromatic fields with 
$\bar{\bf a} = i\omega^{-1} \bar{\bf v} = - (e/m\omega^2) \bar{\bf E}$ and $\bar{\bf A} = -i\omega^{-1} \bar{\bf E}$, we find that the first (kinetic) and the third (coupling) terms in Eq.~(\ref{eq34}) cancel each other, while the remaining second (field) term yields Eq.~(\ref{eq22}) with the electric-field canonical momentum density.

The angular momentum density is calculated similarly to Eqs.~(\ref{eq17}) and (\ref{eq18}), which results in
\begin{eqnarray}
\label{eq35}
{\bf J}_{E} & = & {\bf r}\times {\bf P}_E + {\bf S}_E\, , \\
\label{eq36}
{\bf S}_{E} & = & -mn_0 {\bf v}\times{\bf a} +\frac{1}{4\pi}{\bf E}\times{\bf A} - en_0 {\bf A}\times{\bf a}\,. 
\end{eqnarray}
Again, the first (kinetic) and the third (coupling) terms are mutually cancelled after time averaging for monochromatic fields, and Eq.~(\ref{eq36}) yields the electric-field spin density (\ref{eq23}).

Construction of a dual-symmetric Lagrangian formalism which would yield canonical momentum and spin densities (\ref{eq24}) and (\ref{eq25}) (if it exists) is an important task for future work. 
Such formalism requires representation of the electric and magnetic fields via two vector-potentials with an additional constraint \cite{Bliokh2013NJP,Cameron2013,Cameron2014,Dressel2015}.

%%%%%%%%%%%%%%%%%%%%%%%%%%%%%%
\section{Direct electric current and magnetization associated with the canonical momentum and spin}
%%%%%%%%%%%%%%%%%%%%%%%%%%%%%%
 
As we mentioned above, the canonical momentum and spin of plasma waves are closely related to the Stokes drift and microscopic rotation of electrons. Since electrons are charged particles, these mechanical phenomena generate the corresponding electromagnetic phenomena. Namely, the drift of electrons produces a {\it direct electric current}, while the microscopic rotation generates a {\it magnetization} of the medium. 

The time-averaged Stokes-drift velocity, \red{determined by the difference between the average Lagrangian and Eulerian velocities of a fluid particle}, can be written in a monochromatic wave as \cite{Bremer2017}
\begin{equation}
\label{eq37}
{\bf u} = \frac{1}{2} \, {\rm Re}\!\left[ (\bar{\bf a}^*\!\cdot {\bm \nabla})\bar{\bf v}\right] .
\end{equation}
For longitudinal Langmuir waves (\ref{eq5}), this yields
\begin{equation}
\label{eq38}
{\bf u}_L = \frac{1}{2\omega} \, {\rm Im}\!\left[ \bar{\bf v}^*\!\cdot ({\bm \nabla})\bar{\bf v}\right] ,
\end{equation}
which immediately provides the canonical momentum density $\bar{\bf P}_L = m n_0 {\bf u}_L$, Eq.~(\ref{eq12}).
In the plane-wave case (\ref{PW_L}), ${\bm \nabla} \to i\,{\bf k}$, these equations coincide with the results of Refs.~\cite{Dewar1972,Dodin2015} for the electron drift velocity and canonical wave momentum \red{(see equation for the mean velocity on page 823 of Ref.~\cite{Dewar1972} and Eq.~(22) in Ref.~\cite{Dodin2015})}. For transverse electromagnetic waves (\ref{eq6}), Eq.~(\ref{eq37}) can be written in two equivalent forms:
\begin{eqnarray}
\label{eq39}
{\bf u}_{EM} & = & \frac{1}{2\omega} \, {\rm Im}\!\left[ \bar{\bf v}^*\!\cdot ({\bm \nabla})\bar{\bf v}\right] 
- \frac{\omega_p^2}{8\pi\omega^2 m n_0 c}\, {\rm Re}\!\left( \bar{\bf E}^* \times \bar{\bf H} \right) \nonumber\\
& = & -\frac{1}{4\omega} {\bm \nabla} \times {\rm Im}\!\left( \bar{\bf v}^* \times \bar{\bf v} \right).
\end{eqnarray}
This form does not correspond to the mechanical part of the canonical momentum, Eq.~(\ref{eq24}), suggested by analogy with Eq.~(\ref{eq12}). Moreover, the Stokes drift (\ref{eq39}) vanishes in a plane electromagnetic wave. 
%This provides one more argument in favour of purely-field momentum density (\ref{eq22}).

In all cases, the Stokes drift of electrons produces the direct electric current density
\begin{equation}
\label{eq40}
\bar{\bf j}_S = e n_0 {\bf u} \, .
\end{equation}
The Stokes drift can also manifest itself in various Doppler effects. In particular, it can cause a nonlinear Doppler-type correction to the effective dispersion of plasma waves: $\omega ({\bf k}) \to \omega ({\bf k}) - {\bf k} \cdot {\bf u}$ \cite{Dewar1972,Dodin2015}.

\red{It is worth noticing that the Stokes drift of the medium particles is generated by a {\it ponderomotive force} acting on these particles \cite{Dodin2012,Bliokh2022_Stokes}. To explain this, one needs to consider a large quasi-monochromatic wavepacket, which passes through a given point and locally behaves as a monochromatic wave with a slowly space- and time-varying amplitude. When the wavepacket is far away, the medium particles are motionless. As it approaches the given point, the wavefield intensity grows and the corresponding ponderomotive forces accelerate the medium particles, such that they acquire drift velocity inside the wavepacket. There are two main contributions to the ponderomotive force: (i) the one proportional to the spatial gradient of the kinetic energy density of the medium particles and (ii) the one given by the time derivative of the canonical wave momentum associated with the particles motion \cite{Bliokh2022_Stokes}. It is the  contribution (ii) that generates the Stokes drift velocity. In turn, the electron drift in an electromagnetic wave in plasma caused by the interaction between the oscillating velocity ${\bf v}$ and magnetic field ${\bf H}$, i.e., the time-averaged Lorentz force, is associated with the gradient ponderomotive force (i) and is not related to the Stokes drift and canonical wave momentum.}

The mechanical electron contribution to the spin density is more straightforward and universal \cite{Nakane2018,Jones1973,Shi2019,Bliokh2017NJP,Bliokh2022}. In all cases, the mechanical angular momentum density from the microscopic electron motion is equal to
\begin{equation}
\label{eq41}
\frac{m n_0}{2} \, {\rm Re}\!\left( \bar{\bf a}^*\!\times \bar{\bf v}\right) = \frac{m n_0}{2\omega} \, {\rm Im}\!\left( \bar{\bf v}^*\!\times \bar{\bf v}\right),
\end{equation}
which yields the spin density $\bar{\bf S}_L$, Eq.~(\ref{eq13}), for Langmuir waves and mechanical part of the spin density $\bar{\bf S}_E^{\rm m}$, Eq.~(\ref{eq25}), for electromagnetic waves. The magnetization density produced by the local rotational motion of electrons can be found by multiplying the angular momentum density (\ref{eq41}) with the standard gyromagnetic ratio $e/2mc$:
\begin{equation}
\label{eq42}
\bar{\bf M} = \frac{e n_0}{4\omega c} \, {\rm Im}\!\left( \bar{\bf v}^*\!\times \bar{\bf v}\right) 
= \frac{e\, \omega_p^2}{16\pi m c\, \omega^3} \, {\rm Im}\!\left( \bar{\bf E}^*\!\times \bar{\bf E}\right).
\end{equation}
This result is well known in plasma physics and electrodynamics of continuous media as the {\it inverse Faraday effect} \cite{Bezzerides1977,Mora1978,Pitaevskii1961,LL_ED,Hertel2006}. The corresponding steady magnetic field generated by waves in plasma are observed in experiments via the optical Faraday rotation and other methods \cite{Stamper1971,Stamper1975,DiVergilio1977}.

%%%%%%%%%%%%%%%%%%%%%%%%%%%%%%
%\begin{widetext}
\begin{table*}
\begin{center}
\def\arraystretch{2.5}
\setlength\tabcolsep{8pt}
%\footnotesize
\begin{tabular}{c c c}
\hline \hline
 & Langmuir waves & Electromagnetic waves \\
\hline
Field properties & ${\bf H} = {\bm \nabla}\times {\bf E} = {\bm \nabla}\times {\bf v} = {\bf 0}$ & $n = {\bm \nabla}\cdot {\bf E} = {\bm \nabla} \cdot {\bf v} = {0}$ \\
Dispersion relation & $\omega^2 = \omega_p^2 + \dfrac{\beta n_0}{m}k^2$ & $\omega^2 = \omega_p^2 + c^2 k^2$\\
Energy, $W$ & $mn_0\dfrac{{\bf v}^2}{2} + \beta\dfrac{n^2}{2} + \dfrac{{\bf E}^2}{8\pi}$ & $mn_0\dfrac{{\bf v}^2}{2}  + \dfrac{1}{8\pi}\!\left({\bf E}^2 + {\bf H}^2\right)$ \\
Energy flux, ${\bf U}$ & $\beta n_0 n {\bf v}$ & $\dfrac{c}{4\pi} ({\bf E}\times {\bf H})$ \\
Kinetic momentum, ${\bm{\mathcal P}}$ & $m n {\bf v}$ & $\dfrac{1}{4\pi c} ({\bf E}\times {\bf H})$ \\
Canonical momentum, $\bar{\bf P}$ & $\dfrac{mn_0}{2\omega}\, {\rm Im}\!\left[ \bar{\bf v}^*\!\cdot ({\bm \nabla}) \bar{\bf v}\right]$ & $\dfrac{1}{8\pi\omega} {\rm Im}\!\left[ \bar{\bf E}^*\!\cdot ({\bm \nabla}) \bar{\bf E}\right]$ or 
$\dfrac{1}{16\pi\omega} {\rm Im}\!\left[ \tilde\varepsilon \, \bar{\bf E}^*\!\cdot ({\bm \nabla}) \bar{\bf E} + \bar{\bf H}^*\!\cdot ({\bm \nabla}) \bar{\bf H}\right]$ \\
Spin angular momentum, $\bar{\bf S}$ & $\dfrac{mn_0}{2\omega}\, {\rm Im}\!\left( \bar{\bf v}^*\!\times \bar{\bf v}\right)$ & $\dfrac{1}{8\pi\omega} {\rm Im}\!\left( \bar{\bf E}^*\!\times \bar{\bf E}\right)$ or $\dfrac{1}{16\pi\omega} {\rm Im}\!\left( \tilde\varepsilon \,\bar{\bf E}^*\! \times \bar{\bf E} + \bar{\bf H}^*\! \times \bar{\bf H} \right)$ \\
Helicity, $\bar{\mathfrak{S}}$ & $0$ & $\dfrac{1}{8\pi\omega \sqrt{\varepsilon}}\, {\rm Im}\! \left( {\bf H}^* \cdot {\bf E} \right)$ \\
Plane-wave relations & $\bar{\bm{\mathcal P}} = \bar{\bf P} =\dfrac{\bar{W}}{\omega}\, {\bf k},~~~ 
\bar{\bf S} = {\bf 0}$ & $\bar{\bm{\mathcal P}} = \bar{\bf P} =\dfrac{\bar{W}}{\omega}\, {\bf k},~~~ 
\bar{\bf S} = \sigma\, \dfrac{\bar{W}}{\omega}\, \dfrac{\bf k}{k} = \bar{\mathfrak{S}}\, \dfrac{\bf k}{k} $ \vspace{0.5em} \\
\hline \hline
\end{tabular}
\end{center}
\caption{\red{Summary of the main properties of Langmuir and electromagnetic waves in plasma. All dynamical properties (energy, momentum, spin, etc.) are given in the form of spatial densities. The overbars indicate time-averaged quantities in monochromatic fields. The canonical momentum and spin densities of electromagnetic waves are given in the `standard' (electric-field-based) and dual-symmetric versions. In all cases, the dynamical wave properties satisfy general field-theory relations (\ref{eq7})--(\ref{eq9}).}}
\label{Table}
\end{table*}
%\end{widetext}
%%%%%%%%%%%%%%%%%%%%%%%%%%%%%%

For Langmuir waves, the magnetization (\ref{eq42}) appears in inhomogeneous fields, and it is typically orthogonal to the field intensity gradients and the main propagation direction. The integral magnetization of a localized Langmuir wave vanishes because of the globally-spinless character of longitudinal waves. In turn, for electromagnetic waves, the magnetization (\ref{eq42}) appears even for a cricularly-polarized plane wave, and it is aligned with the wavevector, see Eq.~(\ref{PW_S}). In this case, we can say that electromagnetic waves carry {\it magnetic moment} proportional to the degree of circular polarization (helicity) $\sigma$. The normalized magnetic moment (per one photon, in the $\hbar =1$ units) for an electromagnetic plane wave can be calculated as \cite{Bliokh2017NJP,Bliokh2017PRL}
\begin{equation}
\label{eq43}
{\bm \mu} = \frac{\omega\, \bar{\bf M}}{\bar{W}} = \sigma \frac{e}{2 m c} \frac{\omega_p^2}{\omega^2}\, {\bm \kappa}\,.
\end{equation}
Remarkably, this magnetic moment allows one to derive the electromagnetic wave dispersion in an external magnetic field ${\bf H}_0 = {\rm const}$ in the first-order approximation {\it without} solving the equations of motion for magnetized plasma \cite{Bliokh2018}. Namely, the Zeeman coupling between the magnetic moment (\ref{eq43}) and ${\bf H}_0$ results in the correction to the dispersion relation (\ref{eq6}): 
\begin{equation}
\label{eq44}
\omega({\bf k}) \to \omega({\bf k}) - {\bm \mu} \cdot {\bf H}_0 = \sqrt{\omega_p^2 +c^2 k^2} +\sigma \dfrac{\omega_c}{2}\dfrac{\omega_p^2}{\omega^2} \cos{\theta_0}\,,
\end{equation}
where $\omega_c = |e| H_0/mc$ is the electron cyclotron frequency and $\theta_0$ is the angle between ${\bf k}$ and ${\bf H}_0$. Equation (\ref{eq44}) is the known dispersion relation for electromagnetic waves in magnetized plasma in the first-order approximation in $\omega_c$ \cite{Krall,Akhiezer_book}.

We finally note that the magnetization density (\ref{eq42}) is generally inhomogeneous, and it generates the electric {\it magnetization current} \cite{Herczynski2012}
\begin{equation}
\label{eq45}
\bar{\bf j}_M = c\, {\bm \nabla} \times \bar{\bf M}\,.
\end{equation}
From Eqs.~(\ref{eq37})--(\ref{eq40}), (\ref{eq42}), and (\ref{eq45}) we find that ${\bf j}_M = - {\bf j}_S$ and the total electric current vanishes for transverse electromagnetic waves, while ${\bf j}_M \neq - {\bf j}_S$ for Langmuir waves. In a plane Langmuir wave, ${\bf j}_M = {\bf 0}$, and the direct electric current is given by ${\bf j}_S$.

%%%%%%%%%%%%%%%%%%%%%%%%%%%%%%
\section{Conclusions}
%%%%%%%%%%%%%%%%%%%%%%%%%%%%%%

We have examined the momentum and angular momentum properties of longitudinal (Langmuir) and transverse (electromagnetic) waves in an isotropic nonrelativistic electron plasma {within the hydrodynamical model}. \red{These properties are summarized in Table~\ref{Table}.} We have considered quantities related to the translational and rotational symmetries of the wave fields with respect to the homogeneous medium, i.e., `pseudo'-momentum and angular momentum. There are two types of such quantities: (i) the kinetic (angular) momentum associated with the energy flux and symmetrized (Belinfante) energy-momentum tensor, and (ii) the canonical momentum and angular momentum (including spin) associated with the conserved Noether currents and canonical energy-momentum tensor. The canonical quantities acquire simple and meaningful time-averaged forms for monochromatic fields. \red{Most importantly, it is the canonical momentum and spin densities that are directly related to mechanical transport phenomena, such as the local rotational motion and slow Stokes drift of the electrons in plasma.}

We have found that the momentum and spin properties of Langmuir waves are similar to those of sound waves in fluids or gases. In particular, the canonical momentum and spin densities are determined by the electron velocity field, without explicitly involving the electric field. For electromagnetic waves, the momentum and spin properties are similar to those of optical fields in dispersive media. However, the Abraham-Minkowski controversy does not appear for plasma waves {in the approximation under consideration}, because the Abraham and Minkowski (with the proper dispersive corrections) momenta become equal to each other for plane waves. Instead, there is a dilemma between the standard (using only the electric field) and dual-symmetric (involving both the electric and magnetic fields) forms of the canonical momentum and spin densities. Although the dual-symmetric form looks more appealing and similar to the Brillouin energy density, the microscopic Lagrangian for coupled electrons and electromagnetic fields results in the free-space-like standard forms of the canonical densities for the electric field, without explicitly involving any plasma parameters.  
 
We have also explored important implications of the canonical momentum and spin densities in observable phenomena: the Stokes drift of electrons and the corresponding direct electric current, as well as the microscopic rotational motion of electrons and the corresponding wave-induced magnetization of plasma (the inverse Faraday effect).

These results demonstrate successful application of the general approach to the kinetic and canonical momentum and angular momentum densities, previously developed mostly for optical and acoustic waves, to the complex light-matter-coupled system supporting both longitudinal and transverse modes with different velocities and dispersions. Despite the complexity of the system under consideration, all quantities derived in this work have rather simple and intuitively clear forms. This illuminates fundamental physical properties of waves in plasma and can be useful for analysis of structured plasma waves and their behaviour. 

It should be noticed, however, that our results were derived in the simplest hydrodynamical model of a homogeneous isotropic collisionless plasma with purely longitudinal and transverse modes. In a more general situation including kinetic effects, magnetic fields, inhomogeneities, etc., the situation can be more complicated \cite{Dodin2022}. 

\red{Some phenomena can be readily incorporated in the model under consideration. For example, collisions can be modeled by an effective friction force $-\gamma m {\bf v}$ in the right-hand side of the first Eq.~(\ref{eq1}). For electromagnetic waves, this results in complex permittivity $\varepsilon = 1 - \dfrac{\omega_p^2}{\omega(\omega + i\gamma)}$ \cite{Krall,Maier_book}. From theory of electromagnetic waves in dispersive dissipative media \cite{LL_ED,Dodin2012} and calculations of ponderomotive forces with the friction term \cite{Bliokh2022_Stokes}, one can conclude that the corrections to the energy and momentum densities appear only in the second order in $\gamma/\omega$, i.e., can be neglected in the case of weak dissipation. 

Also, a two-component plasma fluid can be considered by including mobile ions in the equations of motion, which brings about an additional longitudinal ion acoustic mode \cite{Krall, Akhiezer_book}. The approach of this work can be extended to this case straightforwardly.  

In the important case of a magnetized plasma the medium becomes anisotropic and requires a principally different approach. Indeed, the field-theory Belinfante-Rosenfeld construction underlying our approach is valid only for translationally and rotationally invariant (i.e., isotropic) media. Otherwise it is difficult to introduce a meaningful spin density.} 

%%%%%%%%%%%%%%%%%%%%%%%%%%%%%%
\subsection*{Acknowledgements}
%%%%%%%%%%%%%%%%%%%%%%%%%%%%%%
We are grateful to Professor I. Y. Dodin for helpful discussions and relevant remarks.

%\newpage

%\pagebreak

\bibliography{References}

\end{document}